\documentclass{ws-procs9x6}

\usepackage{graphicx}
\usepackage{times,amssymb}
\newcommand{\dd}{ {\mathrm d} }
\newcommand{\ee}{ {\mathrm e} }

\begin{document}

\title{Testing a Possible Way of Geometrization of the Strong Interaction by a Kaluza\,--\,Klein Star\footnote{\uppercase{T}his work was supported by: \uppercase{H}ungarian \uppercase{N}ational \uppercase{R}esearch \uppercase{F}ound (\uppercase{OTKA}) 
\uppercase{NK}106119, \uppercase{K}104260, \uppercase{K}104292 
and the \uppercase{N}ew\uppercase{C}omp\uppercase{S}tar \uppercase{COST} \uppercase{A}ction \uppercase{MP}1304. \uppercase{A}uthor \uppercase{GGB} was partially supported by the \uppercase{J}\'anos \uppercase{B}olyai  \uppercase{R}esearch \uppercase{S}cholarship of the \uppercase{H}ungarian \uppercase{A}cademy of \uppercase{S}ciences.}}

\author{G.G. BARNAF\"OLDI$^{1}$, SZ. KARSAI$^{1,2}$, B. LUK\'ACS$^{1}$, P. P\'OSFAY$^{1,2}$}

\address{$^1$ Wigner RCP of the Hungarian Academy of Sciences, \\
P.O. Box 49, H-1525, Budapest, Hungary \\ E-mail: barnafoldi.gergely@wigner.mta.hu}

\address{$^2$ E\"otv\"os Lor\'and University, \\ P\'azm\'any P\'eter s\'et\'any 1/A, H-1117, Budapest, Hungary}

\maketitle

\abstracts{
Geometrization of the fundamental interactions has been extensively studied during the century. The idea of introducing compactified spatial dimensions originated by Kaluza and Klein. Following their approach, several model were built representing quantum numbers (e.g. charges) as compactified space-time dimensions. Such geometrized theoretical descriptions of the fundamental interactions might lead us to get closer to the unification of the principle theories. 
Here, we apply a $3+1_C+1$ dimensional theory, which contains one extra compactified spatial dimension $1_C$ in connection with the flavour quantum number in Quantum Chromodynamics. Within our model the size of the $1_C$ dimension is proportional to the inverse mass-difference of the first low-mass baryon states. We used this phenomena to apply in a compact star model -- a natural laboratory for testing the theory of strong interaction and the gravitational theory in parallel. 
Our aim is to test the modification of the measurable macroscopical parameters of  a compact Kaluza\,--\,Klein star by varying the size of the compactified extra dimension. Since larger the $R_C$ the smaller the mass difference between the first spokes of the Kaluza\,--\,Klein ladder resulting smaller-mass stars.  
Using the Tolman\,--\,Oppenheimer\,--\,Volkov equation, we investigate the $M$-$R$ diagram and the dependence of the maximum mass of compact stars. Besides testing the validity of our model we compare our results to the existing observational data of pulsar properties for constraints. 
}


\section{Introduction}
\label{sec:intro}

Unification of the two successful theories: gravitational and quantum fields can happen at the Planck's scale in $3+1$ dimensional spacetime, however, these energy-regions are still far from the scope of the recent highest-energy experimental tests. By introducing extra compactified spatial dimension(s) in Kaluza\,--\,Klein theories, related observations might occur even at few-TeV energies\cite{Kaluza:1921,Klein:1926}. Therefore geometrization of quantum fields, where charges are associated with compactified extra dimensions, are under investigations. 

Since, compactified extra dimension(s) or such topological issues have no direct observables in the classical physics cases, thus search for these can start at the known highest energy systems: (i) collisions of nuclei in high-energy particle accelerators or (ii) inside extreme-energetic celestial objects such as compact stars or black holes\cite{Barnafoldi:2007}. 

So far, we have already presented a simplified alternative Kaluza\,--\,Klein-like description for interactions via including a compactified extra dimension ($1_C$), which we could compare to the observables of strangeness-content compact objects. This idea is extended here in comparison to compact objects involving high-mass baryon states with charme and bottomness flavour content.

In this paper we would like to investigate how the maximum mass of a compact object depends on the size of the compactified extra dimension, $R_C$. We test the modification of the measurable macroscopical parameters of a compact Kaluza\,--\,Klein star by varying the size of the compactified extra dimension. We connect the
$R_C$ value to charme- and bottomness-content baryonic matter. Using the Tolman\,--\,Oppenheimer\,--\,Volkov equation, we study the $M$-$R$ diagram and the dependence of the calculated maximum mass of compact stars. Beside testing the validity of our model we compare our results to the existing observational data of known pulsar properties for constraints and suggest new directions for the future theoretical developments.

\section{The Kaluza\,--\,Klein Star in $3+1_C+1$ Dimensional Spacetime}
\label{sec:kkstars}
 
Observed symmetries of the standard 'low-energy' physical world suggest a space-time with $3+1$ macroscopical dimensions, however this can not exclude the existence of compactified extra dimensions at microscopical scales at extreme energies. Due to this fact constructing a generalized theory of compact stars living in a $3+1_C+1$ dimensional Kaluza\,--\,Klein spacetime must behave as a standard $3+1$ dimensional object as well. One can consider a $3+1_C+1$ dimensional space-time, where the particles have enough energy to 
move into the extra compactified space dimension indicated by $1_C$\cite{Barnafoldi:2003,Barnafoldi:2004,Barnafoldi:2010,Lukacs:2003}. 

In the $3+1_C+1$ dimensional space-time, particles can move freely along the extra $x^5$ direction, however we require a periodic boundary condition, which results a Bohr-type quantization condition for $k_5$ momentum component. Generalization of the Heisenberg uncertainty relation induces an uncertainty in the position with the size of $2 \pi R_C$, where $R_C$ is the compactification radius as introduced above. Moreover, motion into the 5\textsuperscript{th} dimension generates an extra mass term appears as 'excited mass', $\widehat{m}$ in the pure 4-dimensional description: 
\begin{equation}
k_5 = \frac{ n \,\,\hbar c}{R_C} \,\,\,\,\,\,\,\ 
\longrightarrow \,\,\,\,\,\,\,\, 
\widehat{m}=\sqrt{m^2+\left(\frac{n \,\,\hbar c}{R_C}\right)^2 }.
\label{quanta}
\end{equation}
Considering a compactified radius $R_C \sim 10^{-12}- 10^{-13}$ cm this extra 'mass' is ${\widehat m} \sim 100$ MeV, which is a familiar quantity in hadron spectroscopy~\cite{Arkhipov:2004}.

Based on eq.~\eqref{quanta} one can introduce a 'pseudo charge' generated by $k_5$: 
\begin{equation}
        {\widehat q} = \pm \ n \cdot \frac{ 2 \hbar \sqrt{G}}{c\,\,R_C} \ \ ,
\label{pseudoq}
\end{equation}
which acts as vector-scalar interaction. We can directly see that ${\widehat q}$ is {\it not} the electric charge, since it is connected to the gravitational constant and the compactified radius. Indeed ${\widehat q} \, ^2 < 16 \pi G m^2 $, where $G$ is the gravitational constant and $m$ is the rest mass of the elementary particle. This idea is supported by Luk\'acs and Pacher\cite{Lukacs:1985}, which paper reviewed the Brans\,--\,Dicke theory\cite{Brans:1961} and pointed out: the generation of extra mass by movement in the $5$\textsuperscript{th} direction is limited due to the charge-to-mass ratio of the observed particles such as e.g. proton. This model helped to rule out the electro-magnetic being of the charge. On the other hand, applying the model for the weak charge, opened a new direction showing the possibility for the unification and geometrization of the weak and gravitational theories. In this picture the extra compactified 5\textsuperscript{th} dimension represents the e.g. hypercharge or other flavour quantum number (strangeness, charm or bottomness).

\section{Theoretical Background of the Model}
\label{sec:einstein}

Describing a compact object one can assume symmetries, which help us to simplify 
both equations and their solution in the $1+3+1_C$ dimensional space-time. The first and
simplest approximation is the {\it spherical symmetry} of the object in the $1+3+1_C$ dimensional
space-time. Such an object is typically {\it static}, since stays in equilibrium with temperature
close to {\it zero}. Assuming perfect fluid for the inner structure is a standard description
including the {\it isotropy} in the standard 3 spatial dimensions. We generalized the 
{\it Killing-symmetries} in the 5D space-time, but an {\it anisotropy} needs to be chosen for
the fluid in the microscopical 5\textsuperscript{th} dimension. These symmetries yield 
to the metric~\footnote{Coordinates $x^i$ with Latin indices ($i=0,1,2,3$ and $5$) follow the
Einstein convention in the $1+3+1_C$ dimensional space-time.}    
\begin{equation}\label{metric}
\dd s^2= \ee^{2\nu} \, \, \dd t^2 - \ee^{2\lambda} \, \, 
\dd r^2 - r^2 \dd {\Omega}^2-\ee^{2\Phi }({\dd x^5})^2 \ ,
\end{equation}
where functions $\nu$, $\lambda$, and $\Phi$ depend on the radius
$r$ only. Here, we denote the usual spherical elementary surface by $\dd \Omega^2$. From this metric, the 5D Einstein equation remains in the usual standard general form:
\begin{equation}
-\gamma \ T_{ik} = R_{ik} - \frac{1}{2} R^l_{\,\,\, l} \ g_{ik}  \ \ .
\label{Einst}
\end{equation}
This includes the applied 5-dimensional energy-momentum tensor containing energy density $\varepsilon$ and pressure components $P$ and $P_5$ in an anisotrop way $P\neq P_5$:  
\begin{equation}
T^{ik}  = {\rm diag} \  ( \varepsilon \, \ee^{2 \nu},  \, P \, \ee^{2 \lambda},
 \, P\, r^2, \, P\,  r^2 \sin^2 \theta, \, P_5 \, \ee^{2 \Phi} ) \ \ .
\label{tik}  \
\end{equation}
Substituting eqs.~(\ref{metric}) and (\ref{tik}) into the 5D Einstein equation 
in eq.~(\ref{Einst}) one obtains the following differential-equation system:
\begin{eqnarray}
{ -\gamma \ \varepsilon } &=& {- \frac{1}{r^2} } + {\ee^{-2 \lambda}} \left[  \Phi'' + \Phi'^2 - \lambda' \Phi'
  + \frac{2 \Phi'}{r} - \frac{2 \lambda'}{r} + \frac{1}{r^2} \right]  \label{einst1}\,  \\
{ -\gamma  P } &=& \frac{1}{r^2} +  \ee^{-2 \lambda} \left[ { - \nu' \Phi' - \frac{2 \Phi'}{r}} - \frac{2 \nu'}{r} - \frac{1}{r^2} \right]  \label{einst2} \\
{ -\gamma  P } &=& \! { \ee^{-2 \lambda} \! \left[\nu' \lambda'-\nu'' -\nu'^2
{ - \Phi'' - \Phi'^2 - \Phi' (\nu'  - \lambda') - \frac{2 \Phi'}{r}}  - \frac{\nu'}{r} + \frac{2 \lambda'}{r} \right]}  \nonumber \\ 
\label{einst3}  \\
{ -\gamma P_5 }  &=&  { \frac{1}{r^2} } + \ee^{-2 \lambda} \left[  -\nu'' -\nu'^2 + \nu' \lambda'
 - \frac{2\nu'}{r}  + \frac{2\lambda'}{r} - \frac{1}{r^2} \right]
  \label{einst4} \ \ 
\end{eqnarray}
where $\gamma = 8 \pi G / c^4$ and $G$ is the gravitational constant.

A typical neutron star has a surface temperature of $T=10^6-10^8$ K, which is much more below the binding energy of nuclei, $T\approx 10^{10}$ K. Thus one can consider these objects as cold nuclear matter, especially inside the fermion star we can 
use the $T = 0$ approximation. The equation of state (EoS) connects the thermodynamical quantities of the 
fluid, which can be e.g. the particle density, $\rho$. Thus, for the matter in a local co-moving frame we can write
\begin{equation}
\varepsilon = \varepsilon(\rho); \,\,\,\,\, \  P=P(\rho);\,\,\,\,\, \  P_5 = P_5(\rho) \ .
\end{equation}
From the appropriate Bianchi identity we obtain 
\begin{equation}
T^{ir}_{\,\,\,\,\,;r} = 0 \ \ \,\,\, \longrightarrow \ \ \,\,\, 
P' = -\nu '(\varepsilon + P) + (P_5-P) \Phi' \ \, . \, \label{bian}
\end{equation}
The influence of the extra dimensional behavior is seen in the latter eq.~\eqref{bian}, where
in parallel to the normal isotropic pressure, \mbox{$P=P_1=P_2=P_3$} and the pressure in the extra compactified dimension, $P_5$ appears too.

The 5-dimensional Einstein equations~(\ref{einst1}-\ref{einst4}) includes two extra variables, compared to the 4-dimensional formalism: $P_5$ and $\Phi$. However, $P_5(\rho)$ is a known function of density and specified by the actual interaction in the matter. Thus $\Phi(r)$ is the {\it only new degree of freedom} determined by eq.~(\ref{einst4}) as detailed in Ref.\cite{Barnafoldi:2003}.

We generalize the equation of state of the non-interacting 5-dimensional fermionic matter using the thermodynamic potential formally, 
\begin{equation}
\widehat{\Omega}=-\widehat{V} \frac{\widehat{g}}{\beta}\int\limits_0^{\infty} \frac{\dd^4 \widehat{k}}{(2\pi)^4}\left[\ln \left(1+\ee^{-\beta(\sqrt{\textbf{k}^2+\widehat{m}}-\widehat{\mu})}\right)+\ln \left(1+\ee^{-\beta(\sqrt{\textbf{k}^2+\widehat{m}}+\widehat{\mu})}\right)\right],
\label{eq:omegapotextr}
\end{equation}
where the total volume in the 5-dimensional spacetime is $\widehat{V}=2\pi R_C \; V$  including the compactification radius, $R_C$ and volume of the 3-dimensional subspace, $V$. The $\widehat{g}=2$ stands for the multiplicity of the $n^0$ neutron. Further physical quantities denoted by  '\, $\widehat{\ } $ \,' are the same as in the usual thermodynamical potential in the $1+3$ dimensional spacetime.

Due to the quantization of the 5-momenta, $ \widehat{k} $, the integral over its 5\textsuperscript{th} component, $k_5$ in eq.~\eqref{eq:omegapotextr} can be rewritten as sum over the excited states, $n=0, 1,...$: 
\begin{equation}
\int\limits_0^\infty \dd^4 \widehat{k}=\int\limits_0^\infty \dd^3 \textbf{k} \, \dd k_5 \;\;\; \longrightarrow \;\;\; \frac{1}{R_C}\sum_{i=\min(n)}^{\max(n)}\int\limits_0^\infty \dd^3 \textbf{k}.
\label{eq:inttosum}
\end{equation}
Substituting this latter eq.~\eqref{eq:inttosum} into eq.~\eqref{eq:omegapotextr}, we can get back the usual thermodynamical potential:
\begin{eqnarray}
\widehat{\Omega}&=&-\widehat{V} \frac{\widehat{g}}{\beta}\sum_{i=\min(n)}^{\max(n)} \int\limits_0^{\infty} \frac{\dd^3 \textbf{k}}{(2\pi)^3} \times \nonumber \\ 
&& \times \left[\ln \left(1+\ee^{-\beta(\sqrt{\textbf{k}^2+\widehat{m}_i}-\widehat{\mu}_i)}\right)+\ln \left(1+\ee^{-\beta(\sqrt{\textbf{k}^2+\widehat{m}_i}+\widehat{\mu}_i)}\right)\right].
\label{eq:omegapotextr2}
\end{eqnarray}
We note, the standard, usual thermodynamical potential, $\Omega=\Omega(m_i, \mu_i, \beta)$ and the $\widehat{\Omega}=\widehat{\Omega}(\widehat{m}_i, \widehat{\mu}_i, \widehat{\beta})$ given by eq.~\eqref{eq:omegapotextr2} looks formally the same, but their variables and sums over the excited states are different. 

\newpage
Thermodynamical variables can be obtained using the usual definitions in the zero temperature approximation, $T=0$, in a two-fermion component model: 
\begin{eqnarray}
\widehat{\epsilon} & = & \frac{\widehat{g}}{(2\pi)^4}\int_0^{\widehat{k_F}}\left .\widehat{\varepsilon}\dd^4\widehat{\textbf{k}}\right|_{T=0}=  \\
& = & \frac{\widehat{g}}{16\pi^3 R_C}\sum_{n=0}^{1}\left[\widehat{\mu}\sqrt{\widehat{\mu}^2-\widehat{m}^2}\left(\widehat{\mu}^2-\frac{1}{2}\widehat{m}^2\right)+\frac{\widehat{m}^4}{2}\ln\left|\frac{\widehat{m}}{\widehat{\mu}+\sqrt{\widehat{\mu}^2-\widehat{m}^2}}\right|\right],\nonumber 
\label{eq:parcepsextr}
\end{eqnarray}
\begin{eqnarray}
\widehat{P} & = & -\frac{1}{2\pi R_C}\left .\frac{\partial\widehat{\Omega}}{\partial V}\right|_{T=0}= \\
 & = & \frac{\widehat{g}}{48\pi^3R_C}\sum_{n=0}^{1}\left[\widehat{\mu}\sqrt{\widehat{\mu}^2-\widehat{m}^2}\left(\widehat{\mu}^2-\frac{5}{2}\widehat{m}^2\right)+\frac{3}{2}\widehat{m}^4\ln\left|\frac{\widehat{m}}{\widehat{\mu}+\sqrt{\widehat{\mu}^2-\widehat{m}^2}}\right|\right]. \nonumber 
\label{eq:parcpextr}
\end{eqnarray}
\begin{eqnarray}
\widehat{P_5} & = & -\frac{1}{2\pi V}\left .\frac{\partial\widehat{\Omega}}{\partial R_C}\right|_{T=0}= \\
 & = & \frac{\widehat{g}}{48\pi^3R_C^3}\left[\widehat{\mu}\sqrt{\widehat{\mu}^2-\bar{m}^2}\left(5\frac{\widehat{\mu}^2}{\bar{m}^2}-11\right)+6\bar{m}^2\ln\left|\frac{\bar{m}}{\widehat{\mu}+\sqrt{\widehat{\mu}^2-\bar{m}^2}}\right|\right].\nonumber 
\end{eqnarray}
It is easy to see, that based on eqns.~\eqref{eq:parcepsextr} and~\eqref{eq:parcpextr} can be connected to the usual description via these expressions: 
$\epsilon=2\pi R_C \widehat{\epsilon}$ and $P=2\pi R_C \widehat{P}$.

\section{A Special 5 Dimensional Solution}

The solution of Einstein equations~(\ref{einst1}-\ref{einst4}) are quite similar in the 
4- and 5-dimensional spacetimes, however there is no analytic solution only numerical. In case of a specially-chosen pressure component $P_5$, there is a unique solution of the eqs.~(\ref{einst1}-\ref{einst4}), especially with the $\dd \Phi / \dd r=0$ constraint. This means the 5\textsuperscript{th} metric component is constant, which choice in eqs.~(\ref{einst1}-\ref{einst3}) 
lead to the Tolman\,--\,Oppenheimer\,--\,Volkov equation\cite{Glendenning:1997} and can be solved separately with the $\Phi=$ const. condition. In addition,    
eq.~(\ref{einst4}) gives $P_5$ as well. 

We note, the extra dimension has its influence on $\varepsilon(\rho)$ and $P(\rho)$~\cite{Kan:2002}, however solution in the 5-dimensional spacetime does not differ formally from the 4 dimensional neutron star solution (except for $P_5$). Furthermore, applying condition $\Phi \neq $ const., 
one must turn to solve Einstein equations directly numerically.

Neglecting the effects of electromagnetic charge, let us start  
with a neutral, single massive fermion (e.g. 'neutron', with mass $m$) as 
elementary building block of an extra dimensional compact star.
Since the minimal nonzero fifth momentum component is given by 
Bohr-type quantization~(\ref{quanta}), then the extra direction 
of the phase space is not populated until the Fermi-momentum 
$k_F < \hbar c/R_C$. However, at the threshold both $k_5 = \pm \hbar c /R_C$ 
states appear. One can represent this as another ('excited') particle 
with excited mass, $\widehat{m}$ and with non-electric 'charge', $\widehat{q}$. We assume this second particle appears in complete chemical equilibrium with the ground-state 'neutron': $\mu_{\widehat{m}} = \mu_{m}$.

Increasing the non-electric charge $\widehat{q}$ i.e. the excited mass, $\widehat{m}$, this recipe can be repeated for every integer $n$, when $k$ exceeds 
a threshold $n\hbar c / R_C$, and these higher excitations are introduced into
the equation of state.

\section{Results and Discussion}
\label{sec:results}

Applying our model, first we calculated the mass-radius relation, $M$-$R$, which is the key macroscopical observable of compact objects, especially, to get constraints for their inner structure. Due to the lack of precise $M(R)$ data\cite{Demorest,Kerkwiwijk:2010,Giullotine}, the maximal mass of the measurable object can give restrictions to the initial conditions and therefore the inner structure as well. Besides the dependence of the $M$-$R$ relation on the $R_C$ we carry out the $M_{max}$, which were found to be also determined by the size of the extra compactified dimension, represented by the compactified radius. 
\begin{figure}[!h]
\centering
\rotatebox{0}{\includegraphics[width=0.80\textwidth]{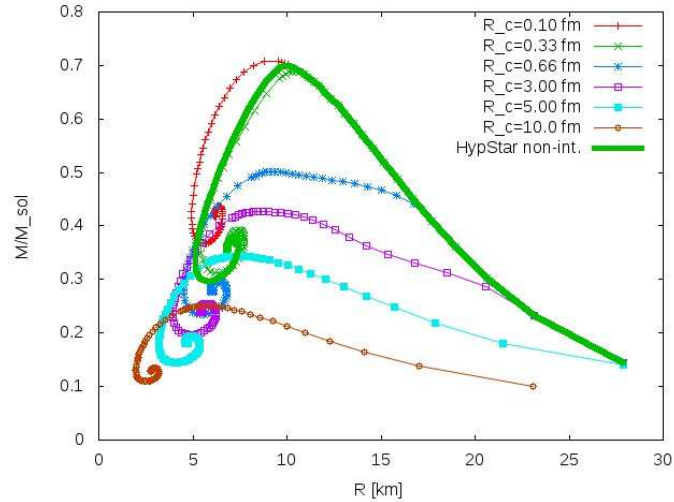}}
\caption{The mass vs. radius relation $M/M_{sol}$-$R$ of compact stars calculated for 1-component
non-interacting fermionic matter in $1+3+1_C$ dimensional space-time for various radii of the extra
compactified dimension, $R_C$ ({\sl lines with dots}). A non-interacting, two-component fermion star
($n^0$ and $\Lambda_s$), is also plotted for comparison ({\sl thick solid green line}).}
\label{fig:1}
\end{figure}


On Figure~\ref{fig:1}, the mass vs. radius relation, $M/M_{sol}$-$R$ of compact stars calculated for 1-component non-interacting fermionic matter in $1+3+1_C$ dimensional space-time for various radii of the extra
compactified dimension, $R_C$ ({\sl lines with dots}). Here $M_{sol}$ is for the Solar-mass unit. We compared this to a reference calculation including a non-interacting, two-component fermion star ($n^0$ and $\Lambda_s$) ({\sl thick solid green line}). To obtain this curve we set the $R_C\approx 0.33$ fm using eq.~\eqref{quanta} with $m_{\Lambda_s}=1115.638$ MeV.

By varying the $R_C$ from 0.1 fm to 10 fm, compact star masses obtained from the extra dimensional model are relatively small: $M$ is between 0.2 and 0.7 $M_{sol}$, with radius and $R$ is in the $2 -25$ km region. As Figure~\ref{fig:1} presents, increasing the size of the compactified extra dimension results smaller-mass objects, which is in agreement with the fact, that mass difference between ground ($m$) and excited ($\widehat{m}$) states gets smaller inducing several excited degree of freedom by the excitation number, $n$. On the other hand, in case of smaller compactification radii, high-mass excited states appear resulting massive compact objects. We note, this simple model does not contain interaction between the fermion states, so relatively light compact stars are obtained. These results can be changed by introducing interaction, as in realistic equation of state parametrizations like e.g. CompOSE\cite{compose}.
\begin{figure}[!h]
\centering
\rotatebox{0}{\includegraphics[width=0.80\textwidth]{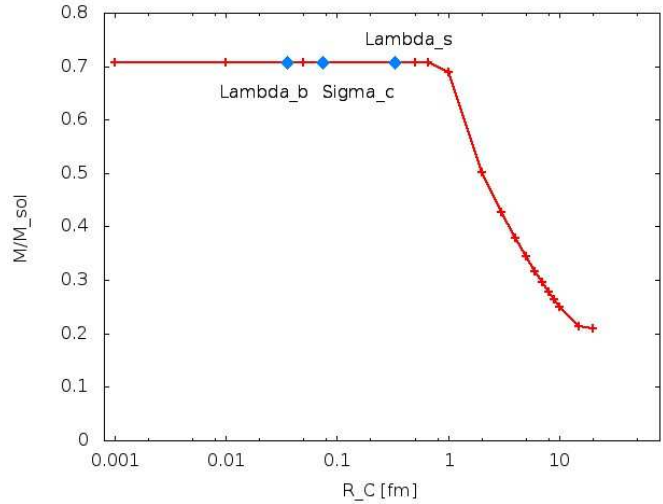}}
\caption{Maximum mass, $M_{max}/M_{sol}$ of compact stars calculated for 1-component non-interacting fermionic matter in $1+3+1_C$ dimensional space-time for various radii of the extra compactified dimension, $R_C$. We highlighted $R_C$ values associated with baryons $\Lambda_s$, $\Sigma_c$, and $\Lambda_b$.}
\label{fig:3}
\end{figure}


Using the obtained $M/M_{sol}$-$R$ curves, we extracted the maximum mass, $M_{max}/M_{sol}$ of compact stars for various radii of the extra compactified dimension, $R_C$ on Figure~\ref{fig:3}. We highlighted some $R_C$ values associated with baryon masses of $\Lambda_s$, $\Sigma_c$, and $\Lambda_b$.

As Figure~\ref{fig:3} presents, the maximum mass is saturating at $M_{max} \approx 0.7 \, M_{sol}$ as $R_C$ gets smaller than $\sim 0.5$ fm. On the other hand the larger the $R_C$ the maximum mass decreasing, which results less-and-less massive stars. 

Increasing the $R_C$ smallers the mass difference between the first spokes of the Kaluza\,--\,Klein ladder. In this case the masses of the excited states are getting closer to each other with small mass gaps between them. This induces several low-mass degrees of freedom which results small-mass stars. For small
$R_C$ values mass difference between the excited states are getting larger. In this case the less, but massive excited particles result the saturation at the highest $M_{max}\approx 0.7 \, M_{sol}$.  

Besides testing the validity of our model we compare our results to the existing observational data of pulsar properties from Ref.~\cite{Demorest,Kerkwiwijk:2010,Giullotine}, which exhibits that our numerical result on $M_{max}/M_{sol}$ is much more below the measured maximal-mass values. This originates from the lack of interaction in our model, what we would like to include into our model in the future.

\section{Summary}

Here we presented a simple model for a possible geometrization of the strongly interacting matter in a Kaluza\,--\,Klein-like spacetime. We applied our approach for modeling the inner structure of a compact star including a compactified extra dimension, with size $R_C$. This was compared to a two-component fermion 
model, where mass difference between ground- and excited-state particle are connected via the compactified radius. We have found that within non-interacting fermionic matter, the $M_{max}\approx 0.7\, M_{sol}$, which is much more below the predictions of either the interacting fermion gas model or the measured maximum mass values $\sim 2.0 M_{sol}$. In order to make our model more realistic, in the future we plan to extend our model with the generalized interaction between fermionic matters, and thus understand more the geometrized excited states in both 
theoretical framework. 



\end{document}